\documentclass[a4paper,aps,twocolumn,nofootinbib]{revtex4}
\RequirePackage[colorlinks,hyperindex]{hyperref}
\RequirePackage[english]{babel}
\RequirePackage[latin1]{inputenc}
\RequirePackage[T1]{fontenc}
\RequirePackage{mathrsfs}
\RequirePackage{amsmath}
\RequirePackage{amssymb}
\RequirePackage{amsbsy}
\RequirePackage{color}
\RequirePackage{bm}
\hypersetup{colorlinks=true,breaklinks=true,urlcolor=blue,linkcolor=red}
\begin{document}
\title{\bf{Spinors in Polar Form}}
\author{Luca Fabbri\footnote{fabbri@dime.unige.it}}
\affiliation{DIME, Sez. Metodi e Modelli Matematici, Universit\`{a} di Genova,\\
Via all'Opera Pia 15, 16145 Genova, ITALY}
\date{\today}
\begin{abstract}
Spinor fields are considered in a generally covariant environment where they can be written in the polar form. The polar form is the one in which spinorial fields are expressed as a module times the exponential of a complex pseudo-phase, and in this form the full spinorial field theory can in turn be expressed by employing only real tensorial quantities. Such a reformulation makes it possible to emphasize properties of the spinorial field theory, and this would enrich our understanding in ways that have never been followed up until this moment.
\end{abstract}
\maketitle
\section{Introduction}
In describing elementary particles such as fermions, it is impossible to underestimate the role of spinorial fields. 

Nevertheless, spinorial fields, defined as vectors in spin space, are intrinsically complex, and this makes it difficult to clearly interpret their structure. For example, of the $4$ complex components, or $8$ real components, constituting a spinor field, it is not immediate to see which are those describing its velocity and spin and which are those accounting for true degrees of freedom in spinor fields.

But precisely because spinorial fields are vectors in spin space, it is possible to attempt a specific type of simplification. In fact, because spinors are columns of $4$ complex components, and each complex component can be written as the product of a module times a unitary phase, it may be possible to perform such a decomposition on all $4$ components, thus reconfiguring the spinor in what should hence be reasonably called polar form. As spinorial fields are objects that transform according to the local complex Lorentz transformation, this procedure would in general spoil the manifest general Lorentz covariance, unless it is carefully implemented. But luckily, there is, as a matter of fact, a way of doing that while respecting the general Lorentz covariance, so writing spinors in polar form can be done in the most general way, and all spinor quantities can therefore be converted into real tensors in general.

In this paper, we present the polar form of spinor fields showing its importance for interpreting the components of the spinor and the isolation of the real scalar degrees of freedom from the components that can be transferred into the frame. In two following sections, we will examine some special situations given by limiting procedures like the non-relativistic approximation. Problems inherent to quantum field theory, and the use of plane waves, will be investigated too. In the final section, we are going to give special attention to the problem of finding exact solutions in the free field configuration. It is important to remark that everything will be treated in the most general cases that are possible. The polar formalism, in transforming the spinor into a clearer object, is what allows us to see features that have never been seen so far. All the original results will be obtained by recovering more efficiently the information that is already contained in the spinorial field theory itself. In no point of the present paper will there be any ad hoc assumption of new physical properties.
\section{General Geometry}
In the first section, we start by giving the general theory of spinorial fields as it is normally presented in common textbooks. Once this will be done, we shall proceed by converting it into the aforementioned polar formalism.
\subsection{Spinor Field}
Because our subject will be spinors, it is necessary to begin by fixing the convention about the Clifford matrices and algebra. They are defined as the matrices $\boldsymbol{\gamma}^{a}$ verifying $\left\{\boldsymbol{\gamma}_{a}\!,\!\boldsymbol{\gamma}_{b}\right\}\!=\!2\eta_{ab}\mathbb{I}$ where $\eta_{ab}$ is the component form of the Minkowskian matrix. This is given according to
\begin{eqnarray}
\eta\!=\!\left(\begin{array}{c|ccc}
\!1&0\!&\!0\!&\!0\!\\
\hline
\!0&-1\!&\!0\!&\!0\!\\
\!0&0\!&\!-1\!&\!0\!\\
\!0&0\!&\!0\!&\!-1\!\\
\end{array}\right)
\end{eqnarray}
such that $\Lambda^{a}_{b}\Lambda^{i}_{j}\eta^{bj}\!\equiv\!\eta^{ai}$ where $\Lambda^{a}_{b}$ is the real representation of the Lorentz transformation. Hence $\left[\boldsymbol{\gamma}_{a}\!,\!\boldsymbol{\gamma}_{b}\right]\!=\!4\boldsymbol{\sigma}_{ab}$ define the $\boldsymbol{\sigma}_{ab}$ matrices that can be proven to verify the commutation relationships of the Lorentz algebra so that with parameters $\theta_{ij}\!=\!-\theta_{ji}$ we write $\boldsymbol{\Lambda}\!=\!\exp{(-\frac{1}{2}\theta_{ab}\boldsymbol{\sigma}^{ab})}$ as the complex representation of the Lorentz transformation in general. Then $\boldsymbol{\Lambda}\boldsymbol{\gamma}^{b}\boldsymbol{\Lambda}^{-1}\Lambda^{a}_{b}\!\equiv\!\boldsymbol{\gamma}^{a}$ hold as identities linking the real to the complex representations of Lorentz transformations. Finally we have that $2i\boldsymbol{\sigma}_{ab}\!=\!\varepsilon_{abcd}\boldsymbol{\pi}\boldsymbol{\sigma}^{cd}$ implicitly define the $\boldsymbol{\pi}$ matrix, which is the matrix usually denoted as a gamma with an index five, but because in the space-time this index has no meaning, we will use a notation with no index. The fact that $\boldsymbol{\pi}$ commutes with all $\boldsymbol{\sigma}_{ab}$ states the reducibility of the complex representation of the Lorentz group. We have that in general
\begin{eqnarray}
&\boldsymbol{\gamma}_{i}\boldsymbol{\gamma}_{j}\boldsymbol{\gamma}_{k}
\!=\!\boldsymbol{\gamma}_{i}\eta_{jk}-\boldsymbol{\gamma}_{j}\eta_{ik}
\!+\!\boldsymbol{\gamma}_{k}\eta_{ij}
\!+\!i\varepsilon_{ijkq}\boldsymbol{\pi}\boldsymbol{\gamma}^{q}
\end{eqnarray}
from which it is possible to prove that
\begin{eqnarray}
&\{\boldsymbol{\gamma}_{a},\boldsymbol{\sigma}_{bc}\}
=i\varepsilon_{abcd}\boldsymbol{\pi}\boldsymbol{\gamma}^{d}\\
&[\boldsymbol{\gamma}_{a},\boldsymbol{\sigma}_{bc}]
=\eta_{ab}\boldsymbol{\gamma}_{c}\!-\!\eta_{ac}\boldsymbol{\gamma}_{b}\label{commgamma}
\end{eqnarray}
and
\begin{eqnarray}
&\{\boldsymbol{\sigma}_{ab},\boldsymbol{\sigma}_{cd}\}
=\frac{1}{2}[(\eta_{ad}\eta_{bc}\!-\!\eta_{ac}\eta_{bd})\mathbb{I}
\!+\!i\varepsilon_{abcd}\boldsymbol{\pi}]\\
&[\boldsymbol{\sigma}_{ab},\boldsymbol{\sigma}_{cd}]
=\eta_{ad}\boldsymbol{\sigma}_{bc}\!-\!\eta_{ac}\boldsymbol{\sigma}_{bd}
\!+\!\eta_{bc}\boldsymbol{\sigma}_{ad}\!-\!\eta_{bd}\boldsymbol{\sigma}_{ac}\label{commsigma}
\end{eqnarray}
are valid as general geometric identities. Because we are working with complex fields, it is always possible to add a unitary phase transformation to the Lorentz transformation thus obtaining $\!\boldsymbol{S}\!\!=\!e^{-i\alpha}\!\boldsymbol{\Lambda}$ as the most general form of what is called spinorial transformation. The spinor field $\psi$ is then defined as what transforms like $\psi\!\rightarrow\!\boldsymbol{S}\psi$ in the most general case. With the Clifford matrices we can see that $\overline{\psi}\!=\!\psi^{\dagger}\boldsymbol{\gamma}^{0}$ transforms like $\overline{\psi}\!\rightarrow\!\overline{\psi}\boldsymbol{S}^{-1}$ and this is what is called adjoint spinor. With a pair of spinor and adjoint spinor we can construct the bi-linear spinor quantities
\begin{eqnarray}
&\Sigma^{ab}\!=\!2\overline{\psi}\boldsymbol{\sigma}^{ab}\boldsymbol{\pi}\psi\\
&M^{ab}\!=\!2i\overline{\psi}\boldsymbol{\sigma}^{ab}\psi
\end{eqnarray}
\begin{eqnarray}
&S^{a}\!=\!\overline{\psi}\boldsymbol{\gamma}^{a}\boldsymbol{\pi}\psi\\
&U^{a}\!=\!\overline{\psi}\boldsymbol{\gamma}^{a}\psi
\end{eqnarray}
\begin{eqnarray}
&\Theta\!=\!i\overline{\psi}\boldsymbol{\pi}\psi\\
&\Phi\!=\!\overline{\psi}\psi
\end{eqnarray}
which are all real tensors. Such a definition of adjoint is unique up to the mere re-definition of these six bi-linear spinor quantities into one another. Nonetheless, these six bi-linear spinor quantities are not independent, since
\begin{eqnarray}
\nonumber
&\psi\overline{\psi}\!\equiv\!\frac{1}{4}\Phi\mathbb{I}
\!+\!\frac{1}{4}U_{a}\boldsymbol{\gamma}^{a}
\!+\!\frac{i}{8}M_{ab}\boldsymbol{\sigma}^{ab}-\\
&-\frac{1}{8}\Sigma_{ab}\boldsymbol{\sigma}^{ab}\boldsymbol{\pi}
\!-\!\frac{1}{4}S_{a}\boldsymbol{\gamma}^{a}\boldsymbol{\pi}
\!-\!\frac{i}{4}\Theta \boldsymbol{\pi}\label{Fierz}
\end{eqnarray}
and then
\begin{eqnarray}
&2\boldsymbol{\sigma}^{\mu\nu}U_{\mu}S_{\nu}\boldsymbol{\pi}\psi\!+\!U^{2}\psi=0\\
&i\Theta S_{\mu}\boldsymbol{\gamma}^{\mu}\psi
\!+\!\Phi S_{\mu}\boldsymbol{\gamma}^{\mu}\boldsymbol{\pi}\psi\!+\!U^{2}\psi=0
\end{eqnarray}
and
\begin{eqnarray}
&\Sigma^{ab}\!=\!-\frac{1}{2}\varepsilon^{abij}M_{ij}\\
&M^{ab}\!=\!\frac{1}{2}\varepsilon^{abij}\Sigma_{ij}
\end{eqnarray}
with
\begin{eqnarray}
&M_{ab}\Phi\!-\!\Sigma_{ab}\Theta\!=\!U^{j}S^{k}\varepsilon_{jkab}\label{A1}\\
&M_{ab}\Theta\!+\!\Sigma_{ab}\Phi\!=\!U_{[a}S_{b]}\label{A2}
\end{eqnarray}
alongside to
\begin{eqnarray}
&M_{ik}U^{i}\!=\!\Theta S_{k}\label{P1}\\
&\Sigma_{ik}U^{i}\!=\!\Phi S_{k}\label{L1}\\
&M_{ik}S^{i}\!=\!\Theta U_{k}\label{P2}\\
&\Sigma_{ik}S^{i}\!=\!\Phi U_{k}\label{L2}
\end{eqnarray}
and with the orthogonality relations
\begin{eqnarray}
&\frac{1}{2}M_{ab}M^{ab}\!=\!-\frac{1}{2}\Sigma_{ab}\Sigma^{ab}\!=\!\Phi^{2}\!-\!\Theta^{2}
\label{norm2}\\
&\frac{1}{2}M_{ab}\Sigma^{ab}\!=\!-2\Theta\Phi
\label{orthogonal2}
\end{eqnarray}
and
\begin{eqnarray}
&U_{a}U^{a}\!=\!-S_{a}S^{a}\!=\!\Theta^{2}\!+\!\Phi^{2}\label{norm1}\\
&U_{a}S^{a}\!=\!0\label{orthogonal1}
\end{eqnarray}
as it is straightforward to demonstrate, and called Fierz identities. We notice that while the definition of the six bi-linear spinor quantities makes up for a symmetric form of the Fierz identities, nevertheless we could drop either $\Sigma^{ab}$ or $M^{ab}$ since they are the dual of each other. We will see in the following that in fact both could be dropped, with simplifications for the Fierz identities. To finish the introduction of all kinematic quantities, we need to say that with the metric we define the symmetric connection as usual with $\Lambda^{\sigma}_{\alpha\nu}$ and with it we define the spin connection according to $\Omega^{a}_{\phantom{a}b\pi}\!=\!\xi^{\nu}_{b}\xi^{a}_{\sigma}(\Lambda^{\sigma}_{\nu\pi}\!-\!\xi^{\sigma}_{i}\partial_{\pi}\xi_{\nu}^{i})$ so that with the gauge potential $qA_{\mu}$ we can finally define
\begin{eqnarray}
&\boldsymbol{\Omega}_{\mu}
=\frac{1}{2}\Omega^{ab}_{\phantom{ab}\mu}\boldsymbol{\sigma}_{ab}
\!+\!iqA_{\mu}\boldsymbol{\mathbb{I}}\label{spinorialconnection}
\end{eqnarray}
called spinorial connection. This is needed to write
\begin{eqnarray}
&\boldsymbol{\nabla}_{\mu}\psi\!=\!\partial_{\mu}\psi
\!+\!\boldsymbol{\Omega}_{\mu}\psi\label{spincovder}
\end{eqnarray}
as spinor covariant derivative. By computing the spinorial covariant derivatives of the gamma matrices and using their constancy we obtain $\boldsymbol{\gamma}^{c}\Omega_{ca\mu}
\!-\![\boldsymbol{\Omega}_{\mu},\boldsymbol{\gamma}_{a}]\!=\!0$ where the general expression 
$\boldsymbol{\Omega}_{\mu}\!=\!a\Omega_{ac\mu}\boldsymbol{\sigma}^{ac}\!+\!\boldsymbol{A}_{\mu}$ might further be plugged in. Once this is done, and using (\ref{commgamma}), we get that $a\!=\!1/2$ and $[\boldsymbol{A}_{\mu},\boldsymbol{\gamma}_{s}]\!=\!0$ identically. Since products of gamma matrices generate the full space of the Clifford matrices then $\boldsymbol{A}_{\mu}$ commutes with every matrix, and this implies that $\boldsymbol{A}_{\mu}\!=\!(iqA_{\mu}\!+\!pC_{\mu})\mathbb{I}$ in general. It is easy to observe that $A_{\mu}$ is the gauge field arising from a complex phase transformation of charge $q$ whereas $C_{\mu}$ is the field arising from conformal transformations $\sigma$ of weight $p$ and this means that with no conformal symmetry the spinor connection given by (\ref{spinorialconnection}) is the most general. And as well known, the commutator of spinorial covariant derivatives can justify the definitions of space-time and gauge tensors
\begin{eqnarray}
&R^{i}_{\phantom{i}j\mu\nu}\!=\!\partial_{\mu}\Omega^{i}_{\phantom{i}j\nu}
\!-\!\partial_{\nu}\Omega^{i}_{\phantom{i}j\mu}
\!+\!\Omega^{i}_{\phantom{i}k\mu}\Omega^{k}_{\phantom{k}j\nu}
\!-\!\Omega^{i}_{\phantom{i}k\nu}\Omega^{k}_{\phantom{k}j\mu}\\
&F_{\mu\nu}\!=\!\partial_{\mu}A_{\nu}\!-\!\partial_{\nu}A_{\mu}
\end{eqnarray}
that is the Riemann curvature and the Maxwell strength.

For the dynamics, we take the spinor field subject to
\begin{eqnarray}
&i\boldsymbol{\gamma}^{\mu}\boldsymbol{\nabla}_{\mu}\psi
\!-\!XW_{\mu}\boldsymbol{\gamma}^{\mu}\boldsymbol{\pi}\psi\!-\!m\psi\!=\!0
\label{D}
\end{eqnarray}
in which $W_{\mu}$ is the axial-vector torsion whereas $X$ is the torsion-spin coupling constant, and this is what is called Dirac equation. If we multiply (\ref{D}) by $\mathbb{I}$, $\boldsymbol{\gamma}^{a}$, $\boldsymbol{\sigma}^{ab}$, $\boldsymbol{\gamma}^{a}\boldsymbol{\pi}$, $\boldsymbol{\pi}$ and then by $\overline{\psi}$ splitting real and imaginary parts gives
\begin{eqnarray}
&\frac{i}{2}(\overline{\psi}\boldsymbol{\gamma}^{\mu}\boldsymbol{\nabla}_{\mu}\psi
\!-\!\boldsymbol{\nabla}_{\mu}\overline{\psi}\boldsymbol{\gamma}^{\mu}\psi)
\!-\!XW_{\sigma}S^{\sigma}\!-\!m\Phi\!=\!0\\
&\nabla_{\mu}U^{\mu}\!=\!0
\end{eqnarray}
\begin{eqnarray}
&\frac{i}{2}(\overline{\psi}\boldsymbol{\gamma}^{\mu}\boldsymbol{\pi}\boldsymbol{\nabla}_{\mu}\psi
\!-\!\boldsymbol{\nabla}_{\mu}\overline{\psi}\boldsymbol{\gamma}^{\mu}\boldsymbol{\pi}\psi)
\!-\!XW_{\sigma}U^{\sigma}\!=\!0\\
&\nabla_{\mu}S^{\mu}\!-\!2m\Theta\!=\!0
\end{eqnarray}
\begin{eqnarray}
\nonumber
&i(\overline{\psi}\boldsymbol{\nabla}^{\alpha}\psi
\!-\!\boldsymbol{\nabla}^{\alpha}\overline{\psi}\psi)
\!-\!\nabla_{\mu}M^{\mu\alpha}-\\
&-XW_{\sigma}M_{\mu\nu}\varepsilon^{\mu\nu\sigma\alpha}\!-\!2mU^{\alpha}\!=\!0
\label{vr}\\
\nonumber
&\nabla_{\alpha}\Phi
\!-\!2(\overline{\psi}\boldsymbol{\sigma}_{\mu\alpha}\!\boldsymbol{\nabla}^{\mu}\psi
\!-\!\boldsymbol{\nabla}^{\mu}\overline{\psi}\boldsymbol{\sigma}_{\mu\alpha}\psi)+\\
&+2X\Theta W_{\alpha}\!=\!0\label{vi}
\end{eqnarray}
\begin{eqnarray}
\nonumber
&\nabla_{\nu}\Theta\!-\!
2i(\overline{\psi}\boldsymbol{\sigma}_{\mu\nu}\boldsymbol{\pi}\boldsymbol{\nabla}^{\mu}\psi\!-\!
\boldsymbol{\nabla}^{\mu}\overline{\psi}\boldsymbol{\sigma}_{\mu\nu}\boldsymbol{\pi}\psi)-\\
&-2X\Phi W_{\nu}\!+\!2mS_{\nu}\!=\!0\label{ar}\\
\nonumber
&(\boldsymbol{\nabla}_{\alpha}\overline{\psi}\boldsymbol{\pi}\psi
\!-\!\overline{\psi}\boldsymbol{\pi}\boldsymbol{\nabla}_{\alpha}\psi)
\!-\!\frac{1}{2}\nabla^{\mu}M^{\rho\sigma}\varepsilon_{\rho\sigma\mu\alpha}+\\
&+2XW^{\mu}M_{\mu\alpha}\!=\!0\label{ai}
\end{eqnarray}
\begin{eqnarray}
\nonumber
&\nabla^{\mu}S^{\rho}\varepsilon_{\mu\rho\alpha\nu}
\!+\!i(\overline{\psi}\boldsymbol{\gamma}_{[\alpha}\!\boldsymbol{\nabla}_{\nu]}\psi
\!-\!\!\boldsymbol{\nabla}_{[\nu}\overline{\psi}\boldsymbol{\gamma}_{\alpha]}\psi)+\\
&+2XW_{[\alpha}S_{\nu]}\!=\!0\\
\nonumber
&\nabla^{[\alpha}U^{\nu]}\!+\!i\varepsilon^{\alpha\nu\mu\rho}
(\overline{\psi}\boldsymbol{\gamma}_{\rho}\boldsymbol{\pi}\!\boldsymbol{\nabla}_{\mu}\psi\!-\!\!
\boldsymbol{\nabla}_{\mu}\overline{\psi}\boldsymbol{\gamma}_{\rho}\boldsymbol{\pi}\psi)-\\
&-2XW_{\sigma}U_{\rho}\varepsilon^{\alpha\nu\sigma\rho}\!-\!2mM^{\alpha\nu}\!=\!0
\end{eqnarray}
as it is direct to see, and called Gordon decompositions.
\subsection{Polar Form}
So far we have presented the general theory of spinorial fields, and it is now our goal to transcribe it into the polar form while respecting its Lorentz covariance. To do so we start by writing each spinor component in polar form, as a product of a module times a unitary phase, but because the spinor transforms according to Lorentz group, the action of such a transformation would inevitably shuffle all these components. In order to circumvent this problem, and save manifest local Lorentz covariance, we move into the special frame in which the only components that remain are the true degrees of freedom, and it will be only on those that we perform the polar decomposition. If we want to make sure that such special frame actually exists, we have to specify that $\Theta$ and $\Phi$ should not be both equal to zero identically. Notice that the case $\Theta\!=\!\Phi\!\equiv\!0$ isolates a class of spinors containing the so-called flagpoles as well as dipoles, themselves respectively containing Majorana spinors and Weyl spinors. Due to the heavy importance that these spinors have in particle physics, it is unwise to blindly dismiss this case as irrelevant. Many mathematical properties have also been recently studied \cite{L, Cavalcanti:2014wia, daSilva:2012wp, Ablamowicz:2014rpa}. But nevertheless, it has also been discussed in \cite{Fabbri:2016msm,Fabbri:2017pwp} that not all features of such spinors are free of problems, and thus we will not consider them in the present treatment. Here we will focus on standard spinors, for which $\Theta$ and $\Phi$ are not identically vanishing at the same time. This makes it possible to see that in (\ref{norm1}) the vector $U^{a}$ is time-like, and therefore there is a special frame in which $U^{a}$ possesses only its time component. Because we also know that such a frame can always be reached by employing only boosts, then $U^{a}$ must be the velocity, and the frame where it has only its temporal component is the frame of rest. This is exactly the special frame we are seeking. In it, (\ref{orthogonal1}) tells that the axial-vector $S^{a}$ has no temporal component, and in general it is always possible to engage two rotations to have $S^{a}$ aligned with some given direction, conventionally chosen to be the third axis. The choice of the third axis means that the two rotations we need can be taken as a rotation around the first axis and a rotation around the second axis, so that the rotation around the third axis is still available. Because the spinor is now an eigen-state of the rotations around the third axis, rotations around the third axis act as one unitary phase. Thus it is possible to employ this rotation to remove a unitary phase. Hence, when written in chiral representation, the most general spinorial field, with components in polar form, and in the frame at rest with spin aligned along the third axis and properly rotated around it, is given by
\begin{eqnarray}
&\!\psi_{\mathrm{polar}}\!=\!\left(\!\begin{tabular}{c}
$\phi e^{\frac{i}{2}\beta}$\\
$0$\\
$\phi e^{-\frac{i}{2}\beta}$\\
$0$
\end{tabular}\!\right)
\end{eqnarray}
in which we can indeed appreciate the polar form of each component, although this form is valid only in the specific frame chosen above. However, the solution is now rather close. Because we may take $\boldsymbol{S}^{-1}$ to be the spinorial transformation that brings $\psi$ into its polar form $\psi_{\mathrm{polar}}$ then it is $\boldsymbol{S}^{-1}\psi\!=\!\psi_{\mathrm{polar}}$ and therefore $\psi\!=\!\boldsymbol{S}\psi_{\mathrm{polar}}$ as clear. Thus the most general spinor field can always be written like
\begin{eqnarray}
&\!\psi\!=\!\boldsymbol{S}\left(\!\begin{tabular}{c}
$\phi e^{\frac{i}{2}\beta}$\\
$0$\\
$\phi e^{-\frac{i}{2}\beta}$\\
$0$
\end{tabular}\!\right)
\end{eqnarray}
for some complex Lorentz transformation $\boldsymbol{S}$ with $\phi$ and $\beta$ called module and Yvon-Takabayashi angle and where we can finally appreciate the polar form of each component and the manifest general Lorentz covariance. Exploiting the fact that $\boldsymbol{\pi}$ commutes with $\boldsymbol{S}$ we can finally write
\begin{eqnarray}
&\!\psi\!=\!\phi e^{-\frac{i}{2}\beta\boldsymbol{\pi}}
\boldsymbol{S}\left(\!\begin{tabular}{c}
$1$\\
$0$\\
$1$\\
$0$
\end{tabular}\!\right)
\label{spinorch}
\end{eqnarray}
and the spinor field is now in \emph{polar form}. When we take the polar form of the spinor field, the antisymmetric tensor bi-linear spinor quantities are given according
\begin{eqnarray}
&\Sigma^{ab}\!=\!2\phi^{2}(\cos{\beta}u^{[a}s^{b]}\!-\!\sin{\beta}u_{j}s_{k}\varepsilon^{jkab})\\
&M^{ab}\!=\!2\phi^{2}(\cos{\beta}u_{j}s_{k}\varepsilon^{jkab}\!+\!\sin{\beta}u^{[a}s^{b]})
\end{eqnarray}
showing that they are not independent quantities, as we had already anticipated, because they can always be constructed in terms of the vector bi-linear spinor quantities
\begin{eqnarray}
&S^{a}\!=\!2\phi^{2}s^{a}\\
&U^{a}\!=\!2\phi^{2}u^{a}
\end{eqnarray}
and the scalar bi-linear spinor quantities
\begin{eqnarray}
&\Theta\!=\!2\phi^{2}\sin{\beta}\\
&\Phi\!=\!2\phi^{2}\cos{\beta}
\end{eqnarray}
which show that module and Yvon-Takabayashi angle are a scalar and a pseudo-scalar and that they amount to the only true degrees of freedom. Fierz identities reduce to
\begin{eqnarray}
&\!\!\!\!\psi\overline{\psi}\!\equiv\!\frac{1}{2}
\phi^{2}[(u_{a}\boldsymbol{\mathbb{I}}\!+\!s_{a}\boldsymbol{\pi})\boldsymbol{\gamma}^{a}
\!\!+\!e^{-i\beta\boldsymbol{\pi}}(\boldsymbol{\mathbb{I}}
\!-\!2u_{a}s_{b}\boldsymbol{\sigma}^{ab}\boldsymbol{\pi})]
\label{F}
\end{eqnarray}
and all others become trivial except for the ones given by
\begin{eqnarray}
&2\boldsymbol{\sigma}^{\mu\nu}u_{\mu}s_{\nu}\boldsymbol{\pi}\psi\!+\!\psi=0\label{aux1}\\
&is_{\mu}\boldsymbol{\gamma}^{\mu}\psi\sin{\beta}
\!+\!s_{\mu}\boldsymbol{\gamma}^{\mu}\boldsymbol{\pi}\psi\cos{\beta}\!+\!\psi=0\label{aux2}
\end{eqnarray}
and
\begin{eqnarray}
&u_{a}u^{a}\!=\!-s_{a}s^{a}\!=\!1\\
&u_{a}s^{a}\!=\!0
\end{eqnarray}
showing that the normalized velocity vector $u^{a}$ and spin axial-vector $s^{a}$ have only three independent components each, and therefore six in total. The advantage of writing spinor fields in polar form is that the $8$ real components are rearranged into the special configuration in which the $2$ real scalar degrees of freedom remain isolated from the $6$ real components that are always transferable into the frame. Consequently, the $2$ real scalar degrees of freedom are given by the module as well as the Yvon-Takabayashi angle, whereas the $6$ real components that can always be transferred into the frame are the $3$ spatial components of the velocity vector and the $3$ spatial components of the spin axial-vector. In fact, the three spatial components of the velocity vector can always be cancelled by the three rapidities while the three spatial components of the spin axial-vector can always be cancelled by the three angles, encoded as the six parameters of the $\boldsymbol{S}$ matrix. The polar form is unique, up to the reversal of the third axis which can always be absorbed as a redefinition of $\boldsymbol{S}$ and up to the discrete transformation $\psi\!\rightarrow\!\boldsymbol{\pi}\psi$ which can always be absorbed as a redefinition of the Yvon-Takabayashi angle with the discrete transformation $\beta\!\rightarrow\!\beta\!+\!\pi$ as clear. It is necessary to notice that because the spinor is a field, the decomposition to the polar form has to be done in terms of frames that are point-dependent, and hence through a transformation that is local. Therefore, we must expect that the Lorentz transformation be gauged. As a matter of fact, in general we can write the following expression
\begin{eqnarray}
&\boldsymbol{S}\partial_{\mu}\boldsymbol{S}^{-1}\!=\!i\partial_{\mu}\alpha\mathbb{I}
\!+\!\frac{1}{2}\partial_{\mu}\theta_{ij}\boldsymbol{\sigma}^{ij}\label{spintrans}
\end{eqnarray}
which can be used for the partial derivative of the spinor field, together with (\ref{spinorialconnection}), in (\ref{spincovder}). We can now define
\begin{eqnarray}
&\partial_{\mu}\alpha\!-\!qA_{\mu}\!\equiv\!P_{\mu}\label{P}\\
&\partial_{\mu}\theta_{ij}\!-\!\Omega_{ij\mu}\!\equiv\!R_{ij\mu}\label{R}
\end{eqnarray}
which can be proven to be tensors and invariant under a gauge transformation simultaneously, and with which
\begin{eqnarray}
&\!\!\!\!\!\!\!\!\boldsymbol{\nabla}_{\mu}\psi\!=\!(-\frac{i}{2}\nabla_{\mu}\beta\boldsymbol{\pi}
\!+\!\nabla_{\mu}\ln{\phi}\mathbb{I}
\!-\!iP_{\mu}\mathbb{I}\!-\!\frac{1}{2}R_{ij\mu}\boldsymbol{\sigma}^{ij})\psi
\label{decspinder}
\end{eqnarray}
is the spinorial covariant derivative. We also have
\begin{eqnarray}
&\nabla_{\mu}s_{i}\!=\!R_{ji\mu}s^{j}\label{ds}\\
&\nabla_{\mu}u_{i}\!=\!R_{ji\mu}u^{j}\label{du}
\end{eqnarray}
are general geometric identities. As we said above, when we write the spinor field in its polar form, the spinor field is reconfigured so that its degrees of freedom are isolated from the components transferable into gauge and frames, with the transfer done through the phase $\alpha$ and the parameters $\theta_{ij}$ which thus contain only information about gauge and frames. While the phase and parameters that are added to the gauge potential and spin connection do not alter the information content of these last, in such a combination all non-covariant characters cancel, so that (\ref{P}, \ref{R}) contain the same information of gauge potential and spin connection, while being gauge invariant as well as Lorentz covariant. Therefore $P_{\mu}$ and $R_{ij\mu}$ can be said to be a \emph{gauge-invariant vector momentum} and a \emph{tensorial connection}. Then taking the commutator of the spinorial covariant derivatives of the spinor field, or the covariant derivatives of the velocity or spin, we obtain that
\begin{eqnarray}
\!\!\!\!&qF_{\mu\nu}\!=\!-(\nabla_{\mu}P_{\nu}\!-\!\nabla_{\nu}P_{\mu})\label{Maxwell}\\
&\!\!\!\!\!\!\!\!R^{i}_{\phantom{i}j\mu\nu}\!=\!-(\nabla_{\mu}R^{i}_{\phantom{i}j\nu}
\!-\!\!\nabla_{\nu}R^{i}_{\phantom{i}j\mu}
\!\!+\!R^{i}_{\phantom{i}k\mu}R^{k}_{\phantom{k}j\nu}
\!-\!R^{i}_{\phantom{i}k\nu}R^{k}_{\phantom{k}j\mu})\label{Riemann}
\end{eqnarray}
in terms of the Maxwell strength and Riemann curvature, and so they encode electrodynamic and gravitational information, filtering out all information about gauge and frames. But we recall that the gauge-invariant vector momentum and the tensorial connection encode information about electrodynamics and gravity as well as about gauge and frames. If it were possible to find non-zero solutions of equations (\ref{Maxwell}, \ref{Riemann}) after setting the Maxwell strength and Riemann curvature to zero, they would represent the gauge-invariant vector momentum and the tensorial connection encoding information related to gauge and frames solely. This information would be non-trivial as encoded by non-zero objects but it would be covariant as encoded in tensors. In this situation the resulting gauge-invariant vector momentum and tensorial connection would hence describe a gauge-invariant vector potential and a covariant inertial acceleration. This might appear strange, but the angular momentum displays an analogous character, because in a macroscopic system it could always be calculated with respect to the reference system in which the particle rests, where it vanishes, but for some microscopic quantum systems it includes the spin, which can not be zero. So a quantity that is non-covariant macroscopically but which may turn out to be covariant microscopically for quantum systems should not really be regarded as too strange after all. But nevertheless, we remark that, even if we can no longer say that the inertial accelerations are those accelerations that are not covariant, we could still define them as the accelerations that do not have sources, much in the same way in which the vector potential may have an action, even in a situation where it does not have any strength, like in the Aharonov-Bohm effect. We shall see in a while that it is in fact possible to find a situation in which $P_{\mu}\!\neq\!0$ and $R_{ij\mu}\!\neq\!0$ with $F_{\mu\nu}\!=\!0$ and $R_{ij\mu\nu}\!=\!0$ thus describing a gauge-invariant vector potential and a covariant inertial acceleration as we discussed above.

For the Dirac equations (\ref{D}), plugging into the Gordon decompositions (\ref{vi}, \ref{ar}) the polar form allows us to write the manifestly covariant polar form of Dirac equations
\begin{eqnarray}
\nonumber
&\frac{1}{2}\nabla_{\alpha}\ln{\phi^{2}}\cos{\beta}
\!-\!(\frac{1}{2}\nabla_{\alpha}\beta\!-\!XW_{\alpha})\sin{\beta}+\\
\nonumber
&+P^{\mu}(u^{\rho}s^{\sigma}\varepsilon_{\rho\sigma\mu\alpha}\cos{\beta}
\!+\!u_{[\mu}s_{\alpha]}\sin{\beta})+\\
&+\frac{1}{2}R_{\alpha\mu}^{\phantom{\alpha\mu}\mu}\cos{\beta}
\!+\!\frac{1}{4}R^{\rho\sigma\mu}\varepsilon_{\rho\sigma\mu\alpha}\sin{\beta}\!=\!0\\
\nonumber
&\frac{1}{2}\nabla_{\nu}\ln{\phi^{2}}\sin{\beta}
\!+\!(\frac{1}{2}\nabla_{\nu}\beta\!-\!XW_{\nu})\cos{\beta}+\\
\nonumber
&+P^{\mu}(u^{\rho}s^{\sigma}\varepsilon_{\rho\sigma\mu\nu}\sin{\beta}\!-\!u_{[\mu}s_{\nu]}\cos{\beta})-\\
&-\frac{1}{4}R^{\rho\sigma\mu}\varepsilon_{\rho\sigma\mu\nu}\cos{\beta}
\!+\!\frac{1}{2}R_{\nu\mu}^{\phantom{\nu\mu}\mu}\sin{\beta}\!+\!ms_{\nu}\!=\!0
\end{eqnarray}
which can be diagonalized into
\begin{eqnarray}
\nonumber
&\frac{1}{4}\varepsilon_{\mu\alpha\nu\iota}R^{\alpha\nu\iota}
\!\!-\!\!P^{\iota}u_{[\iota}s_{\mu]}\!-\!\!XW_{\mu}+\\
&+\nabla_{\mu}\beta/2\!+\!s_{\mu}m\!\cos{\beta}\!=\!0\\
\nonumber
&\frac{1}{2}R_{\mu a}^{\phantom{\mu a}a}
\!-\!P^{\rho}u^{\nu}s^{\alpha}\varepsilon_{\mu\rho\nu\alpha}+\\
&+s_{\mu}m\sin{\beta}\!+\!\nabla_{\mu}\ln{\phi}\!=\!0
\end{eqnarray}
in general. Conversely, from these and (\ref{aux1}, \ref{aux2}) we have
\begin{eqnarray}
\nonumber
&i\boldsymbol{\gamma}^{\mu}\boldsymbol{\nabla}_{\mu}\psi
\!-\!XW_{\sigma}\boldsymbol{\gamma}^{\sigma}\boldsymbol{\pi}\psi\!-\!m\psi=\\
\nonumber
&=[P^{\rho}(i\boldsymbol{\gamma}^{\mu}u^{\nu}s^{\alpha}\varepsilon_{\mu\rho\nu\alpha}
\!+\!u_{[\rho}s_{\mu]}\boldsymbol{\gamma}^{\mu}\boldsymbol{\pi}
\!+\!\boldsymbol{\gamma}_{\rho})-\\
&-m(is_{\mu}\boldsymbol{\gamma}^{\mu}\sin{\beta}
\!+\!s_{\mu}\boldsymbol{\gamma}^{\mu}\boldsymbol{\pi}\cos{\beta}\!+\!\mathbb{I})]\psi\!=\!0
\end{eqnarray}
showing that the initial Dirac equations are valid. So we proved that the Dirac equations (\ref{D}) are equivalent to
\begin{eqnarray}
&\!\!\!\!B_{\mu}\!-\!2P^{\iota}u_{[\iota}s_{\mu]}\!+\!(\nabla\beta\!-\!2XW)_{\mu}
\!+\!2s_{\mu}m\cos{\beta}\!=\!0\label{dep1}\\
&\!\!\!\!R_{\mu}\!-\!2P^{\rho}u^{\nu}s^{\alpha}\varepsilon_{\mu\rho\nu\alpha}\!+\!2s_{\mu}m\sin{\beta}
\!+\!\nabla_{\mu}\ln{\phi^{2}}\!=\!0\label{dep2}
\end{eqnarray}
with $R_{\mu a}^{\phantom{\mu a}a}\!=\!R_{\mu}$ and $\frac{1}{2}\varepsilon_{\mu\alpha\nu\iota}R^{\alpha\nu\iota}\!=\!B_{\mu}$ as it has been quite extensively discussed in \cite{Fabbri:2016laz, Fabbri:2018crr}. The spinor equations (\ref{D}) consist of $4$ complex equations, that is $8$ real equations, which are as many as the $2$ vectorial equations given by the (\ref{dep1}, \ref{dep2}) above. Such a pair of vector equations specify all space-time derivatives of both degrees of freedom given by module and Yvon-Takabayashi angle. It is very important to say that the discrete transformation given above as $\beta\!\!\rightarrow\!\!\beta+\pi$ requires that $m\!\rightarrow\!-m$ be implemented too in order for (\ref{dep1}, \ref{dep2}) to be invariant. Notice that the Yvon-Takabayashi angle, in its being the phase difference between chiral projections, has to be related to the mass term, which makes up for a basic interaction between the two parts. Also notice that because the interplay of such chiral parts is reminiscent of internal dynamics then also the presence of the spin axial-vector is expected. We will make this comment more solid a little later in the work.
\section{Special Structure}
At some point in the previous section we talked about the fact that the tensorial connection might describe the covariant inertial acceleration and therefore an effect due to no source, if we were able to find a non-zero tensorial connection which is solution of the zero Riemann curvature equation. The similar consideration holds for finding non-zero gauge-invariant vector momentum as solution of the zero Maxwell strength equation. Next we will provide such an example. But we will not stop there, as we shall use this example to find more general spinor solutions of the non-interacting Dirac differential field equations.
\subsection{Background}
The zero Riemann curvature equation $R^{i}_{\phantom{i}j\mu\nu}\!=\!0$ must be solved for a non-zero tensorial connection $R_{ijk}$ but it may be difficult in general. So will make some simplifying assumption, and the first is to pick spherical coordinates with metric given by
\begin{eqnarray}
&g_{tt}\!=\!1\\
&g_{rr}\!=\!-1\\
&g_{\theta\theta}\!=\!-r^{2}\\
&g_{\varphi\varphi}\!=\!-r^{2}|\!\sin{\theta}|^{2}
\end{eqnarray}
giving connection
\begin{eqnarray}
&\Lambda^{\theta}_{\theta r}\!=\!\frac{1}{r}\\
&\Lambda^{\varphi}_{\varphi r}\!=\!\frac{1}{r}\\
&\Lambda^{r}_{\theta\theta}\!=\!-r\\
&\Lambda^{r}_{\varphi\varphi}\!=\!-r|\!\sin{\theta}|^{2}\\
&\Lambda^{\varphi}_{\varphi\theta}\!=\!\cot{\theta}\\
&\Lambda^{\theta}_{\varphi\varphi}\!=\!-\cos{\theta}\sin{\theta}
\end{eqnarray}
and zero Riemann curvature as wanted. Nevertheless, a truly simplifying hypothesis is that of exploiting relationships (\ref{ds}, \ref{du}) to find some $R_{ijk}$ which only later will be restricted to be compatible with the $R^{i}_{\phantom{i}j\mu\nu}\!=\!0$ condition we want to implement. Exploiting (\ref{ds}, \ref{du}) means that a set of assumptions must be taken for the vectors $u_{k}$ and $s_{k}$ in a careful manner. Therefore, to begin, we might demand that $u_{k}$ have temporal and azimuthal components solely, so that due to normalization, they are taken as
\begin{eqnarray}
&u_{t}\!=\!\cosh{\alpha}\label{u1}\\
&u_{\varphi}\!=\!r\sin{\theta}\sinh{\alpha}\label{u2}
\end{eqnarray}
with $\alpha\!=\!\alpha(r,\theta)$ generic function. Orthogonality between vectors $u_{k}$ and $s_{k}$ indicates that we may take $s_{k}$ having the radial and elevational components solely, so that due to normalization, we can take
\begin{eqnarray}
&s_{r}\!=\!\cos{\gamma}\label{s1}\\
&s_{\theta}\!=\!r\sin{\gamma}\label{s2}
\end{eqnarray}
with $\gamma\!=\!\gamma(r,\theta)$ another generic function. Now relations (\ref{ds},\ref{du}) can be solved for $R_{ijk}$ and we obtain that
\begin{eqnarray}
&R_{t\varphi\varphi}\!=\!R_{\varphi tt}\!=\!R_{r\theta\varphi}\!=\!R_{r\theta t}\!=\!0
\end{eqnarray}
as well as
\begin{eqnarray}
&r\sin{\theta}\partial_{\theta}\alpha\!=\!R_{t\varphi\theta}\\
&r\sin{\theta}\partial_{r}\alpha\!=\!R_{t\varphi r}\\
&-r(1\!+\!\partial_{\theta}\gamma)\!=\!R_{r\theta\theta}\\
&r\partial_{r}\gamma\!=\!R_{\theta rr}
\end{eqnarray}
linking the derivatives of the two above functions to four of the components of the $R_{ijk}$ tensor and
\begin{eqnarray}
&\!rR_{rt\varphi}\!=\!R_{t\theta\varphi}\tan{\gamma}\\
&\!\!r\sin{\theta}R_{t\theta\varphi}\!=\!(R_{\varphi\theta\varphi}
\!-\!r^{2}\cos{\theta}\sin{\theta})\tanh{\alpha}\\
&\!\!\!\!(R_{\varphi\theta\varphi}\!-\!r^{2}\sin{\theta}\cos{\theta})\tan{\gamma}\!=\!
r(R_{r\varphi\varphi}\!+\!r|\!\sin{\theta}|^{2})\\
&\!\!(R_{r\varphi\varphi}\!+\!r|\!\sin{\theta}|^{2})\tanh{\alpha}
\!=\!r\sin{\theta}R_{rt\varphi}
\end{eqnarray}
as well as
\begin{eqnarray}
&rR_{rtt}\!=\!R_{t\theta t}\tan{\gamma}\\
&r\sin{\theta}R_{t\theta t}\!=\!R_{\varphi\theta t}\tanh{\alpha}\\
&R_{\varphi\theta t}\tan{\gamma}\!=\!rR_{r\varphi t}\\
&R_{r\varphi t}\tanh{\alpha}\!=\!r\sin{\theta}R_{rtt}
\end{eqnarray}
and
\begin{eqnarray}
&rR_{rtr}\!=\!R_{t\theta r}\tan{\gamma}\\
&r\sin{\theta}R_{t\theta r}\!=\!R_{\varphi\theta r}\tanh{\alpha}\\
&R_{\varphi\theta r}\tan{\gamma}\!=\!rR_{r\varphi r}\\
&R_{r\varphi r}\tanh{\alpha}\!=\!r\sin{\theta}R_{rtr}
\end{eqnarray}
with
\begin{eqnarray}
&rR_{rt\theta}\!=\!R_{t\theta\theta}\tan{\gamma}\\
&r\sin{\theta}R_{t\theta\theta}\!=\!R_{\varphi\theta\theta}\tanh{\alpha}\\
&R_{\varphi\theta\theta}\tan{\gamma}\!=\!rR_{r\varphi\theta}\\
&R_{r\varphi\theta}\tanh{\alpha}\!=\!r\sin{\theta}R_{rt\theta}
\end{eqnarray}
grouped in four blocks of four relations. In each of these blocks the four components are mutually related, but different blocks are independent on one another. Thus it is possible to set an entire block to zero while leaving different from zero all others. So as another hypothesis and educated guess we can finally choose the system having
\begin{eqnarray}
&R_{trr}\!=\!R_{t\theta r}\!=\!R_{\varphi rr}\!=\!R_{\varphi\theta r}\!=\!0\\
&R_{rt\theta}\!=\!R_{t\theta\theta}\!=\!R_{r\varphi\theta}\!=\!R_{\varphi\theta\theta}\!=\!0\\
&R_{t\theta\varphi}\!=\!R_{tr\varphi}\!=\!0
\end{eqnarray}
with
\begin{eqnarray}
&R_{r\varphi\varphi}\!=\!-r|\!\sin{\theta}|^{2}\\
&R_{\theta\varphi\varphi}\!=\!-r^{2}\cos{\theta}\sin{\theta}
\end{eqnarray}
and
\begin{eqnarray}
&R_{rtt}\!=\!-2\varepsilon\sinh{\alpha}\sin{\gamma}\\
&R_{\varphi rt}\!=\!2\varepsilon r\sin{\theta}\cosh{\alpha}\sin{\gamma}\\
&R_{\theta tt}\!=\!2\varepsilon r\sinh{\alpha}\cos{\gamma}\\
&R_{\varphi\theta t}\!=\!-2\varepsilon r^{2}\sin{\theta}\cosh{\alpha}\cos{\gamma}
\end{eqnarray}
with $\varepsilon$ being a generic function a priori. However it will turn out to be a constant after imposing the vanishing of the Riemann curvature we wanted to have. Remark that the above choice (\ref{u1}, \ref{u2}), (\ref{s1}, \ref{s2}) gives the tetrads
\begin{eqnarray}
&\!\!\!\!e^{0}_{t}\!=\!\cosh{\alpha}\ \ \ \ 
e^{2}_{t}\!=\!\sinh{\alpha}\\
&\!\!\!\!e^{1}_{r}\!=\!\sin{\gamma}\ \ \ \ 
e^{3}_{r}\!=\!-\cos{\gamma}\\
&\!\!\!\!e^{1}_{\theta}\!=\!-r\cos{\gamma}\ \ \ \ 
e^{3}_{\theta}\!=\!-r\sin{\gamma}\\
&\!\!\!\!e^{0}_{\varphi}\!=\!r\sin{\theta}\sinh{\alpha}\ \ \ \ 
e^{2}_{\varphi}\!=\!r\sin{\theta}\cosh{\alpha}
\end{eqnarray}
and dual tetrads
\begin{eqnarray}
&\!\!\!\!e_{0}^{t}\!=\!\cosh{\alpha}\ \ \ \ e_{2}^{t}\!=\!-\sinh{\alpha}\\
&\!\!\!\!e_{1}^{r}\!=\!\sin{\gamma}\ \ \ \ e_{3}^{r}\!=\!-\cos{\gamma}\\
&\!\!\!\!e_{1}^{\theta}\!=\!-\frac{1}{r}\cos{\gamma}\ \ \ \ 
e_{3}^{\theta}\!=\!-\frac{1}{r}\sin{\gamma}\\
&\!\!\!\!e_{0}^{\varphi}\!=\!-\frac{1}{r\sin{\theta}}\sinh{\alpha}\ \ \ \ 
e_{2}^{\varphi}\!=\!\frac{1}{r\sin{\theta}}\cosh{\alpha}
\end{eqnarray}
from which the spin connection is
\begin{eqnarray}
&\Omega_{02r}\!=\!-\partial_{r}\alpha\\
&\Omega_{13r}\!=\!-\partial_{r}(\theta\!+\!\gamma)\\
&\Omega_{02\theta}\!=\!-\partial_{\theta}\alpha\\
&\Omega_{13\theta}\!=\!-\partial_{\theta}(\theta\!+\!\gamma)\\
&\Omega_{01\varphi}\!=\!-\cos{(\theta\!+\!\gamma)}\sinh{\alpha}\\ 
&\Omega_{03\varphi}\!=\!-\sin{(\theta\!+\!\gamma)}\sinh{\alpha}\\
&\Omega_{23\varphi}\!=\!\sin{(\theta\!+\!\gamma)}\cosh{\alpha}\\ 
&\Omega_{12\varphi}\!=\!-\cos{(\theta\!+\!\gamma)}\cosh{\alpha}
\end{eqnarray}
which is obviously a non-trivial spin connection although it gives a zero Riemann curvature. With the tetrads we can write the tensorial connection in tetradic formalism in the first two indices so that by means of (\ref{R}) and the spin connection we get $\partial_{t}\theta_{12}\!=\!-2\varepsilon$ alone. Consequently, we have established the existence of a non-zero tensorial connection solution of zero Riemann curvature equations.

The zero Maxwell strength equation $F_{\mu\nu}\!=\!0$ has to be solved for a non-zero gauge-invariant vector momentum $P_{\mu}$ but this is rather easy. In fact $P_{\mu}\!=\!\partial_{\mu}\alpha$ with $P_{t}\!=\!E$ and $E$ a generic constant will already suffice. Therefore, (\ref{P}) will give $\alpha\!=\!Et$ alone. Consequently, we have again found a non-zero gauge-invariant vector momentum that is a general solution of zero Maxwell strength equations.

We will next see how this background can be used to find more general solutions of the Dirac equations.
\subsection{Matter}
We have given an example of non-zero gauge-invariant vector momentum and tensorial connection that have no Maxwell strength and Riemann curvature. Next we shall employ them to find more general spinor solutions of the non-interacting Dirac equations. In fact, a zero Riemann curvature means flat space-time and thus no gravity and a zero Maxwell strength means no electrodynamics. They can be complemented by the condition of no torsion thus ensuring to neglect all of the interactions within the Dirac equations. Just the same, the fact that we have non-zero gauge-invariant vector momentum and tensorial connection means that some effect must be present even though we are dealing with free Dirac equations. A background of this type is trivial but still non-negligible, and in it the field equations given by (\ref{dep1}, \ref{dep2}) become expressed as
\begin{eqnarray}
\nonumber
&\!\!\!\!\partial_{\theta}\alpha\!-\!2(\varepsilon\!+\!E)r\cosh{\alpha}\cos{\gamma}+\\
&+r\partial_{r}\beta\!+\!2mr\cos{\beta}\cos{\gamma}\!=\!0\label{deps1}\\
\nonumber
&\!\!\!\!r\partial_{r}\alpha\!+\!2(\varepsilon\!+\!E)r\cosh{\alpha}\sin{\gamma}-\\
&-\partial_{\theta}\beta\!-\!2mr\cos{\beta}\sin{\gamma}\!=\!0\label{deps2}\\
\nonumber
&\!\!\partial_{\theta}\gamma\!-\!2(\varepsilon\!+\!E)r\sin{\gamma}\sinh{\alpha}+\\
&+2mr\cos{\gamma}\sin{\beta}
\!+\!r\partial_{r}\ln{(\phi^{2}r^{2}\sin{\theta})}\!=\!0\label{deps3}\\
\nonumber
&\!\!-r\partial_{r}\gamma\!+\!2(\varepsilon\!+\!E)r\cos{\gamma}\sinh{\alpha}+\\
&+2mr\sin{\gamma}\sin{\beta}
\!+\!\partial_{\theta}\ln{(\phi^{2}r^{2}\sin{\theta})}\!=\!0\label{deps4}
\end{eqnarray}
which have to be solved in some special case.

With this example we conclude the presentation of the general theory, and now we turn to study specific cases.
\section{Relativistic Effect}

\subsection{Zitterbewegung}
Up to now we have investigated the general theory and some exact solutions. It is now time to consider specific situations which, despite being special cases, might reveal interesting information. The first condition that we will consider is the non-relativistic approximation. To study this condition we start by specifying that after requiring the non-relativistic limit, it will turn out that time is not a dimension anymore and boosts are no longer a possible transformation. When this is done, we essentially remain with a genuinely $3$-dimensional theory. A spinor field in $3$ dimensions has $2$ complex, and thus $4$ real, components, and it is subject to $3$ transformations, that is the three rotations. By re-doing the polar analysis on such a spinor field we would find that it will basically have one degree of freedom, which is recognized as the module. Because we have that in $4$ dimensions a spinor field has two degrees of freedom, the module and Yvon-Takabayashi angle, while in $3$ dimensions spinor fields have one degree of freedom, the module, then the process of having the general theory reduced in non-relativistic regime has to account for the vanishing of the Yvon-Takabayashi angle. This must take place beside the already implemented condition given by the vanishing the three space components of the velocity vector. As a consequence, such a non-relativistic approximation will be expressed by conditions $\beta\!\rightarrow\!0$ and $\vec{u}\!\rightarrow\!0$ in the most general circumstance that is indeed possible.

When the $4$-dimensional spinor field is written in polar form we are naturally equipped to do the non-relativistic approximation. Indeed requiring zero Yvon-Takabayashi angle is immediate and requiring zero space velocity is implemented by boosting into the rest frame. When this last condition is combined with the always possible alignment of the spin along the third axis we are merely asking that $\boldsymbol{S}$ be the identity. Then the spinor is given by
\begin{eqnarray}
&\!\psi\!=\!\phi\left(\!\begin{tabular}{c}
$1$\\
$0$\\
$1$\\
$0$
\end{tabular}\!\right)
\end{eqnarray}
in chiral representation or
\begin{eqnarray}
&\!\psi\!=\!\phi \sqrt{2}
\!\left(\!\begin{tabular}{c}
$1$\\
$0$\\
$0$\\
$0$
\end{tabular}\!\right)
\end{eqnarray}
in standard representation. So the non-relativistic regime can be implemented by a single condition asking that the spinor field written in standard representation has lower component that is vanishing. To see the reciprocal, it is necessary to write the spinor field in polar form explicitly, and a way to do that is according to the expression
\begin{eqnarray}
&\!\!\!\!\psi\!=\!\phi\sqrt{\frac{2}{\gamma+1}}e^{-i\alpha}
e^{-\frac{i}{2}\beta\boldsymbol{\pi}}\!\!
\left(\!\begin{tabular}{c}
$\left(\frac{\gamma+1}{2}\boldsymbol{\mathbb{I}}\!-\!
\gamma\vec{v}\!\cdot\!\vec{\frac{\boldsymbol{\sigma}}{2}}\right)\!\xi$\\
$\left(\frac{\gamma+1}{2}\boldsymbol{\mathbb{I}}\!+\!
\gamma\vec{v}\!\cdot\!\vec{\frac{\boldsymbol{\sigma}}{2}}\right)\!\xi$
\end{tabular}\right)
\label{spinorexch}
\end{eqnarray}
in chiral representation or
\begin{eqnarray}
&\!\!\!\!\!\!\!\!\psi\!=\!\phi\sqrt{2}\sqrt{\frac{2}{\gamma+1}}e^{-i\alpha}\!
\!\left(\!\begin{tabular}{c}
$\!\left(\cos{\frac{\beta}{2}}\frac{\gamma+1}{2}\mathbb{I}
\!-\!i\sin{\frac{\beta}{2}}\gamma\vec{v}
\!\cdot\!\vec{\frac{\boldsymbol{\sigma}}{2}}\right)\!\xi$\\
$\!\left(\cos{\frac{\beta}{2}}\gamma\vec{v}
\!\cdot\!\vec{\frac{\boldsymbol{\sigma}}{2}}
\!-\!i\sin{\frac{\beta}{2}}\frac{\gamma+1}{2}\mathbb{I}\right)\!\xi$
\end{tabular}\right)
\label{spinorexst}
\end{eqnarray}
in standard representation, where $\gamma\!=\!1/\!\sqrt{1\!-\!v^{2}}$ in terms of the spatial velocity $\gamma\vec{v}\!=\!\vec{u}$ with $\xi$ such that $\xi^{\dagger}\xi\!=\!1$ is a generic semi-spinor and $\alpha$ a unitary phase. This explicit form shows that the vanishing of the lower component is given, for whatever direction of the semi-spinor $\xi$, by
\begin{eqnarray}
&\cos{\frac{\beta}{2}}\vec{u}\!\cdot\!\vec{\boldsymbol{\sigma}}
\!-\!i\sin{\frac{\beta}{2}}(\gamma+1)\mathbb{I}\!=\!0
\end{eqnarray}
and because $\mathbb{I}$ and all $\vec{\boldsymbol{\sigma}}$ are linearly independent then we must have $\beta\!=\!0$ and $\vec{u}\!=\!0$ as well. So the non-relativistic approximation is equivalently implemented by requiring the single condition given by the vanishing of the lower component of spinors in standard representation. Notice that this is why normally the lower component is called small component. Also, remark that for a discrete transformation $\beta\!\!\rightarrow\!\!\beta\!+\!\pi$ the lower component will convert into the upper component having the role of small component.

There is yet another way for the non-relativistic regime to be expressed, and it consists in having zero space part of the momentum. To study this condition we need find a way to have the momentum written in terms of velocity and Yvon-Takabayashi angle. For this purpose, consider the Dirac equations in polar form (\ref{dep1}, \ref{dep2}) and contract the first by $u^{\mu}$ and $s^{\mu}$ getting
\begin{eqnarray}
&\!\!\!\!P\!\cdot\!s\!+\!\frac{1}{2}(\nabla\beta\!-\!2XW\!+\!B)\!\cdot\!u\!=\!0\label{Ps}\\
&\!\!\!\!\!\!\!\!P\!\cdot\!u\!+\!\frac{1}{2}(\nabla\beta\!-\!2XW\!+\!B)\!\cdot\!s
\!-\!m\cos{\beta}\!=\!0\label{Pu}
\end{eqnarray}
and the second by $u^{\nu}s^{\alpha}\varepsilon_{\nu\alpha\mu\rho}$ getting
\begin{eqnarray}
\nonumber
&P^{\rho}\!+\!P\!\cdot\!s s^{\rho}\!-\!P\!\cdot\!u u^{\rho}+\\
&+\frac{1}{2}(\nabla\ln{\phi^{2}}\!+\!R)_{\mu}s_{\alpha}u_{\nu}
\varepsilon^{\mu\alpha\nu\rho}\!=\!0
\end{eqnarray}
and then substitute in the last one all occurrences of $P\!\cdot\!s$ and $P\!\cdot\!u$ in terms of the other two expressions getting
\begin{eqnarray}
\nonumber
&\!\!\!\!P^{\rho}\!=\!m\cos{\beta}u^{\rho}
\!-\!\frac{1}{2}(\nabla\beta\!-\!2XW\!+\!B)\!\cdot\!s u^{\rho}+\\
\nonumber
&+\frac{1}{2}(\nabla\beta\!-\!2XW\!+\!B)\!\cdot\!u s^{\rho}-\\
&-\frac{1}{2}(\nabla\ln{\phi^{2}}\!+\!R)_{\mu}s_{\alpha}u_{\nu}
\varepsilon^{\mu\alpha\nu\rho}
\end{eqnarray}
in general. In doing this we demonstrated that the Dirac equations in polar form (\ref{dep1}, \ref{dep2}) imply the expression
\begin{eqnarray}
&P^{\mu}\!=\!Mu^{\mu}\!+\!Y\!\cdot\!u s^{\mu}\!+\!Z_{k}s_{j}u_{i}\varepsilon^{kji\mu}
\label{momentum}
\end{eqnarray}
where $m\cos{\beta}\!-\!Y\!\cdot\!s\!=\!M$ with $\frac{1}{2}(\nabla\beta\!-\!2XW\!+\!B)_{k}\!=\!Y_{k}$ and $-\frac{1}{2}(\nabla\ln{\phi^{2}}\!+\!R)_{k}\!=\!Z_{k}$ to have a more compact expression of the momentum \cite{Fabbri:2019tad}. We notice that in such a form the momentum has a component along the velocity, but also a component along the spin and a component orthogonal to velocity and spin. It is possible also to have this form inverted, and to do so we can first dot it into $Z_{a}$ to get
\begin{eqnarray}
&P\!\cdot\!Z\!+\!P\!\cdot\!s Z\!\cdot\!s\!=\!MZ\!\cdot\!u
\end{eqnarray}
and then dot it into $Z^{i}s^{j}\varepsilon_{ijka}$ to get
\begin{eqnarray}
\nonumber
&P^{a}Z^{i}s^{j}\varepsilon_{ijka}\!=\!MZ^{i}s^{j}u^{a}\varepsilon_{ijka}+\\
&+(Z^{2}\!+\!|Z\!\cdot\!s|^{2})u_{k}\!-\!Z\!\cdot\!u (Z_{k}\!+\!Z\!\cdot\!s s_{k})
\end{eqnarray}
and after having the first substituted into the second, and the result plugged back into (\ref{momentum}), we get
\begin{eqnarray}
\nonumber
&u^{k}\!=\!(1\!+\!\zeta^{2}\!+\!|\zeta\!\cdot\!s|^{2})^{-1}[\eta^{ka}+\\
\nonumber
&+s^{a}s^{k}(1\!+\!|\zeta\!\cdot\!s|^{2})\!+\!\zeta^{a}\zeta^{k}+\\
&+(s^{a}\zeta^{k}\!+\!\zeta^{a}s^{k})\zeta\!\cdot\!s\!+\!\zeta_{i}s_{j}\varepsilon^{ijka}]P_{a}/M
\label{eq}
\end{eqnarray}
with $\zeta_{k}\!=\!Z_{k}/M$ and giving the expression of the velocity in terms of the momentum. In the following this expression will become important. Back to the expression of the momentum in terms of the velocity and also involving the Yvon-Takabayashi angle, we have for the non-relativistic limit that the momentum is eventually given by
\begin{eqnarray}
&P^{0}\!=\!m\!-\!(X\vec{W}\!-\!\frac{1}{2}\vec{B})\!\cdot\!\vec{s}\\
&\vec{P}\!=\!-(XW^{0}\!-\!\frac{1}{2}B^{0})\vec{s}
\!-\!\frac{1}{2}(\vec{\nabla}\ln{\phi^{2}}\!-\!\vec{R})\!\times\!\vec{s}
\end{eqnarray}
showing that the energy does not reduce to the mass and the space momentum does not vanish. This case can be obtained only when we neglect the spin. Thus it is solely when we neglect the spin that $P^{\mu}\!=\!m\cos{\beta}u^{\mu}$ and hence the non-relativistic limit given by $P_{a}\!\rightarrow\!(m,\vec{0})$ coincides with the non-relativistic regime $\beta\!\rightarrow\!0$ and $\vec{u}\!\rightarrow\!0$ according to the general procedure that we have discussed here.

Therefore, the condition of non-relativistic limit can be implemented by asking that $\beta\!\rightarrow\!0$ and $\vec{u}\!\rightarrow\!0$ or equivalently by asking that in standard representation the small component vanishes, and in case the spin is negligible by asking $P_{a}\!\!\rightarrow\!\!(m,\vec{0})$ to hold. Notice that the first two forms are equivalent, but the form $\beta\!\rightarrow\!0$ and $\vec{u}\!\rightarrow\!0$ allows us to interpret the Yvon-Takabayashi angle. In fact, if on the one hand we cannot get non-relativistic limit even in rest frame if the Yvon-Takabayashi angle does not vanish, on the other hand we will always have some dynamical character that remains even at rest if an internal dynamics is present. Thus, we interpret the Yvon-Takabayashi angle as what encodes information about the internal dynamics of the spinor. When the small component is not zero it gives rise to an effect called zitterbewegung, which is the jittering motion of relativistic quantum particles. Thus, a more precise identification of the internal dynamics would simply be that of zitterbewegung. The last form we have discussed for the non-relativistic limit is the one given in terms of the momentum and despite being the form that is commonly used nevertheless its validity is subsequent to the condition of neglecting the spin. This means that once again we have relativistic effects even if the momentum is small if spin contributions are considered. Thus, the phenomenology attributed to relativistic effects for a quantum particle can be transposed onto the presence of non-zero Yvon-Takabayashi angles or small components of a spinor, representing internal dynamics or zitterbewegung, and subsequently also to the spin contributions.

We will use some of the material presented in this section in the following treatment of quantum fields.
\section{Quantum Correction}

\subsection{Anomaly}
In the previous section we analyzed the non-relativistic conditions, so that when the most general spinor field was compared to its non-relativistic approximation it became possible to isolate the relativistic effects. Because a fully relativistic quantum dynamics naturally leads to the concept of quantum fields, we next turn to a general study of quantum field theory. Because in QFT one of the basic assumptions is to perform calculations employing plane waves, we will see how the most general spinor would be written as one fixed plane wave plus specific field corrections. The plane wave form is implemented by requiring
\begin{eqnarray}
&i\boldsymbol{\nabla}_{\mu}\psi\!=\!P_{\mu}\psi
\end{eqnarray}
in general. Upon comparison with the most general form given by expression (\ref{decspinder}) we obtain that
\begin{eqnarray}
&(\nabla_{\mu}\ln{\phi}\mathbb{I}
\!-\!\frac{i}{2}\nabla_{\mu}\beta\boldsymbol{\pi}
\!-\!\frac{1}{2}R_{ij\mu}\boldsymbol{\sigma}^{ij})\psi=0
\end{eqnarray}
for any spinor field and because $\boldsymbol{\sigma}^{ij}$, $\mathbb{I}$ and $\boldsymbol{\pi}$ are linearly independent then $R_{ij\mu}=0$ with $\beta$ and $\phi$ constant. Since a constant pseudo-scalar has to vanish then we finally get that $R_{ij\mu}=0$ with $\beta\!=\!0$ and $\nabla_{\nu}\phi\!=\!0$ are the conditions that implement the reduction to standard QFT. Thus, in QFT the completeness relationships (\ref{F}) are
\begin{eqnarray}
&\!\!\!\!\psi\overline{\psi}\!\equiv\!\frac{1}{2}\phi^{2}
(u_{a}\boldsymbol{\gamma}^{a}\!+\!\boldsymbol{\mathbb{I}}
\!-\!s_{a}\boldsymbol{\gamma}^{a}\boldsymbol{\pi}
\!-\!2u_{a}s_{b}\boldsymbol{\sigma}^{ab}\boldsymbol{\pi})
\end{eqnarray}
with spin and velocity, then by using (\ref{eq}) reduced as
\begin{eqnarray}
&u^{k}\!=\!(P^{k}\!+\!P\!\cdot\!ss^{k})/(m\!+\!XW\!\cdot\!s)
\end{eqnarray}
we can write the velocity in terms of the momentum, so that the completeness relationships can be written with spin and momentum, as conserved quantities. The Dirac equations (\ref{dep1}, \ref{dep2}) reduce to the simpler
\begin{eqnarray}
&P^{\iota}u_{[\iota}s_{\mu]}\!+\!XW_{\mu}\!-\!s_{\mu}m\!=\!0\\
&P^{\rho}u^{\nu}s^{\alpha}\varepsilon_{\mu\rho\nu\alpha}\!=\!0
\end{eqnarray}
as a link between momentum and torsion and a constraint on the momentum, so after some algebraic manipulation
\begin{eqnarray}
&XW_{\alpha}u^{[\alpha}s^{\mu]}\!+\!P^{\mu}\!-\!mu^{\mu}\!=\!0\\
&XW^{\alpha}(s_{\alpha}s_{\mu}\!-\!u_{\alpha}u_{\mu}\!+\!g_{\alpha\mu})\!=\!0
\end{eqnarray}
expressing the momentum in terms of torsion and a constraint on torsion and telling us that while in QFT torsion can be important nevertheless the zero-torsion condition can be taken to simplify the Dirac equations. With no torsion, the Dirac equations are equivalent to $P^{\mu}\!=\!mu^{\mu}$ in general. The completeness relationships are
\begin{eqnarray}
&\psi\overline{\psi}\!\equiv\!\frac{\phi^{2}}{2m}
[m\boldsymbol{\mathbb{I}}\!+\!P_{a}\boldsymbol{\gamma}^{a}
\!-\!ms_{a}\boldsymbol{\gamma}^{a}\boldsymbol{\pi}
\!-\!2P_{a}s_{b}\boldsymbol{\sigma}^{ab}\boldsymbol{\pi}]
\end{eqnarray}
which are those of QFT. In fact by neglecting the spin we eventually reduce to the form $\psi\overline{\psi}\!=\!(P_{k}\boldsymbol{\gamma}^{k}\!+\!m\boldsymbol{\mathbb{I}})(\phi^{2}/2m)$ as the common expressions for the spin-sum used in QFT.

What we found is that the usual formalism of QFT is recovered by the conditions $\nabla_{\nu}\phi\!=\!0$ and $\beta\!=\!0$ alongside to $R_{ij\mu}=0$ and $XW_{\mu}\!=\!0$ when also the spin is averaged out so to be neglected. Any relaxation of these conditions means an enlargement of QFT worth studying. The most general case is given by the completeness relationships
\begin{eqnarray}
&\!\!\!\!\!\!\!\!\psi\overline{\psi}\!\equiv\!\frac{1}{2}
\phi^{2}[(u_{a}\boldsymbol{\mathbb{I}}\!+\!s_{a}\boldsymbol{\pi})\boldsymbol{\gamma}^{a}
\!\!+\!e^{-i\beta\boldsymbol{\pi}}(\boldsymbol{\mathbb{I}}
\!-\!2u_{a}s_{b}\boldsymbol{\sigma}^{ab}\boldsymbol{\pi})]
\end{eqnarray}
where the velocity can be expressed as
\begin{eqnarray}
\nonumber
&u^{k}\!=\!(1\!+\!\zeta^{2}\!+\!|\zeta\!\cdot\!s|^{2})^{-1}[\eta^{ka}+\\
\nonumber
&+s^{a}s^{k}(1\!+\!|\zeta\!\cdot\!s|^{2})\!+\!\zeta^{a}\zeta^{k}+\\
&+(s^{a}\zeta^{k}\!+\!\zeta^{a}s^{k})\zeta\!\cdot\!s\!+\!\zeta_{i}s_{j}\varepsilon^{ijka}]P_{a}/M
\label{eqexplicit}
\end{eqnarray}
in terms of spin and momentum as conserved quantities but with $\zeta_{k}$ and $M$ given by
\begin{eqnarray}
&\zeta_{k}\!=\!Z_{k}/M\\
&M\!=\!m\cos{\beta}\!-\!Y\!\cdot\!s
\end{eqnarray}
in terms of $Z_{k}$ and $Y_{k}$ defined by
\begin{eqnarray}
&Z_{k}\!=\!-\frac{1}{2}(\nabla\ln{\phi^{2}}\!+\!R)_{k}\\
&Y_{k}\!=\!\frac{1}{2}(\nabla\beta\!-\!2XW\!+\!B)_{k}
\end{eqnarray}
with torsion, tensorial connection, Yvon-Takabayashi angle and module. We also have the constraints
\begin{eqnarray}
&Z\!\cdot\!s\!+\!m\sin{\beta}\!=\!0\label{depm}\\
&Z\!\cdot\!u\!=\!0\label{depZ}\\
&Y_{\mu}u_{\nu}s_{\rho}\varepsilon^{\mu\nu\rho\sigma}\!=\!0\label{depY}
\end{eqnarray}
which are four independent equations since the last does not have any projections along the directions of velocity and spin, and therefore (\ref{depY}), (\ref{depZ}) and (\ref{depm}) together with (\ref{eqexplicit}) make a total of eight independent equations exactly accounting for the full set of the Dirac equations.

It is very intriguing to notice that the condition $\beta\!=\!0$ is common to both the non-relativistic limit and the plane wave structure of QFT. Also, it is interesting to observe that commonly we neglect or average the spin away again both in the non-relativistic approximation and for plane wave solutions in QFT. While the former condition need be implemented, the latter condition only appears to be a working hypothesis. Nonetheless, some procedure has to be employed in order to conceal the intrinsic structure of elementary particles in both situations. This is comprehensible, because we are used to consider all elementary particles as point-like objects. But even if this might be true to a good extent, there is no real reason for that.

Accounting for extended fields with intrinsic structure would be more comprehensive, and this is what we intend to do in the development of the following section.
\section{Solution}
The past sections were devoted to justify why it is more comprehensive to perform a study of extended fields displaying intrinsic structures. In this section we are going to present an example of exact solution. Such a solution will be chosen to have an internal structure of trivial type, represented by the condition $\beta\!=\!0$ and taking $\gamma\!=\!\pi/2$ for simplicity. In this way taking $E\!=\!m$ gives
\begin{eqnarray}
&\partial_{\theta}\alpha\!=\!0\\
&\partial_{\theta}\ln{(\phi^{2}r^{2}\sin{\theta})}\!=\!0\\
&\partial_{r}\alpha\!+\!2(\varepsilon\!+\!m)\cosh{\alpha}\!-\!2m\!=\!0\\
&-2(\varepsilon\!+\!m)\sinh{\alpha}\!+\!\partial_{r}\ln{(\phi^{2}r^{2}\sin{\theta})}\!=\!0
\end{eqnarray}
having set $\phi^{2}r^{2}\sin{\theta}\!\equiv\!F\!=\!F(r)$ and with $\alpha\!=\!\alpha(r)$ in the most general case. A solution is obtained according to
\begin{eqnarray}
&\!\!\tanh{\frac{\alpha}{2}}=-\sqrt{\frac{\varepsilon}{(\varepsilon+2m)}}
\tan{\left[r\sqrt{\varepsilon(\varepsilon\!+\!2m)}\right]}\\
&\!\!\!\!F\!=\!A\left[2(\varepsilon\!+\!m)
\left|\cos{\left[r\sqrt{\varepsilon(\varepsilon\!+\!2m)}\right]}\right|^{2}
\!-\!\varepsilon\right]
\end{eqnarray}
for $\varepsilon\!>\!0$ and with $A$ integration constant provided that $(\varepsilon\!+\!m)\cosh{\alpha}\!\neq\!m$ while for $(\varepsilon\!+\!m)\cosh{\alpha}\!=\!m$ we choose
\begin{eqnarray}
&\tanh{\frac{\alpha}{2}}\!=\!-\sqrt{\frac{-\varepsilon}{2m+\varepsilon}}\\
&F\!=\!Be^{-2r\sqrt{-\varepsilon(\varepsilon+2m)}}
\end{eqnarray}
for $-2m\!<\!\varepsilon\!<\!\!0$ and with $B$ integration constant. Because a positive $\varepsilon$ gives rise to an oscillatory behaviour whereas a negative $\varepsilon$ can give rise to an exponentially decreasing behaviour, we might interpret it as some potential energy creating a potential barrier when positive and a potential well when negative, or equivalently we might think at the tensorial connection as a sort of tension that can be either repulsive or attractive \cite{Fabbri:2019kfr}. It is also remarkable that the condition $|\varepsilon|\!<\!2m$ tells that the potential well cannot be deeper than at most two times the mass of the particle.

The former solution is well-behaved at the origin but does not converge, while the latter solution would have a cusp at the origin but an exponential damping at infinity, suggesting that some form of junction may form a single solution with good properties everywhere. This junction is formed by cutting both solutions at a given radius $R$ and piecing them together, requiring both continuity and smoothness on the boundary. For the inner solution, we can set $\varepsilon\!=\!|\varepsilon|\!=\!\varepsilon_{1}$ while, for the outer solution, we can set $\varepsilon\!=\!-|\varepsilon|\!=\!-\varepsilon_{2}$ obtaining the condition of continuity
\begin{eqnarray}
\nonumber
&A\left[2(\varepsilon_{1}\!+\!m)
\left|\cos{\left[R\sqrt{\varepsilon_{1}(\varepsilon_{1}\!+\!2m)}\right]}\right|^{2}\!-\!\varepsilon_{1}\right]=\\
&=Be^{-2R\sqrt{\varepsilon_{2}(2m-\varepsilon_{2})}}
\end{eqnarray}
and smoothness
\begin{eqnarray}
\nonumber
&A(\varepsilon_{1}\!+\!m)\sqrt{\varepsilon_{1}(\varepsilon_{1}\!+\!2m)}
\sin{\left[2R\sqrt{\varepsilon_{1}(\varepsilon_{1}\!+\!2m)}\right]}=\\
&=B\sqrt{\varepsilon_{2}(2m-\varepsilon_{2})}e^{-2R\sqrt{\varepsilon_{2}(2m-\varepsilon_{2})}}
\end{eqnarray}
which can be worked out to give
\begin{eqnarray}
&\frac{2(\varepsilon_{1}+m)
\left|\cos{\left[R\sqrt{\varepsilon_{1}(\varepsilon_{1}+2m)}\right]}\right|^{2}-\varepsilon_{1}}
{(\varepsilon_{1}+m)\sin{\left[2R\sqrt{\varepsilon_{1}(\varepsilon_{1}+2m)}\right]}}
\!=\!\sqrt{\frac{\varepsilon_{1}(2m+\varepsilon_{1})}{\varepsilon_{2}(2m-\varepsilon_{2})}}
\label{w}
\end{eqnarray}
plus another relationship determining the size of the $A/B$ constant. Nonetheless, for us (\ref{w}) is the most important relationship because it gives a constraint that can be read as a discretization condition for the mass of the particle.

Equation (\ref{w}) is a terrifying relationship to solve, but it might still be informative. In fact, in the case of a very large $\varepsilon_{1}$ we can approximate (\ref{w}) to a relationship that is solved for the mass in terms of the expression
\begin{eqnarray}
2Rm\!=\!R\varepsilon_{2}
\!+\!\frac{1}{R\varepsilon_{2}}\left|R\varepsilon_{1}\tan{(2R\varepsilon_{1})}\right|^{2}
\label{m}
\end{eqnarray}
which tells that there is an infinity of possible values of the mass constant. As we have no control over $R$, $\varepsilon_{1}$ and $\varepsilon_{2}$ the specific value of the mass in (\ref{m}) is free, but still it is determined by the other constants of the problem.

This situation is totally analogous to what we observe to happen in non-relativistic cases, where the junction of two solutions determines discretization conditions of the energy. In relativistic cases it happens exactly the same, but with the energy replaced by the mass. It is important to remark also that if we were to calculate the tensorial connection we would find it different from zero, meaning that such a solution cannot be quantized in terms of the common prescriptions used for field quantization.

This solution is what one would obtain in presence of a positive tension $\varepsilon_{1}$ close to the origin which relaxes into an overall negative tension $\varepsilon_{2}$ as we move away from the origin. Such potential mimics what we would have in the presence of an overall attractive gravitational field which would turn into a repulsive gravitational field as we move close to the origin of the matter distribution. As strange as this situation may seem, it is precisely what happens when the field equations of gravity are written in presence of propagating torsion sourced by the spinor field \cite{Fabbri:2017rjf}.

The interpretation we have given of this situation is a straightforward one. When spinorial fields source torsion, the spin-torsion interaction effectively acts between chiral components as an attractive force. Attractive potentials give rise to negative contributions that for large densities might become dominant, reverting the energy density to negative, thus inverting the sign of the curvature and so the gravitational response. This feature may appear to be counterintuitive, but it is merely the result of combining a theory in which gravity is sourced by an energy density to a theory in which that energy density might turn out to be negative in some conditions, which is precisely what happens when Einstein gravitation is sourced by a Dirac spinor in presence of a propagating massive torsion field. 

Notice however that the analysis leading to (\ref{m}) is not to be taken too seriously because it strongly depends on the specific solution we considered, determined with the constraints $\gamma\!=\!\pi/2$ and $\beta\!=\!0$ which were taken to remove the intrinsic structure so to simplify the treatment. However, a realistic solution must consider internal structure, that is the information related to the spin content and to the Yvon-Takabayashi angle in the most general case.

We leave the search of such a solution and the investigations of some of its properties to following works.
\section{Conclusion}
In this paper, we have presented, in the most general mathematical setting, the Dirac spinor and its field equations in polar form, the one that isolates real scalar degrees of freedom from all components that can be transferred into the frame and there combined with the underlying geometrical features therefore giving rise to the gauge-invariant vector momentum and the tensorial connection. We have studied the non-relativistic approximation and the plane-wave structure of quantum fields for which the polar form helps to highlight a few important properties, such as the consequences that come from the dismissal of the information related to spin. And finally, we have considered a potential mimicking real situations but with easier structure, actually finding a solution and discussing it in terms of conditions that may give rise to a discrete mass spectrum for the material distribution.

These results are of a very general interest, but leaving aside the specific content of information that these results have, our main purpose was more formal, and it consisted in presenting a way in which by employing the polar form it is possible to assess in the clearest manner the physical meaning of spinors, allowing a more efficient study of the properties that pertain to such field theory and which are overlooked when employing usual methods.

\end{document}